# Static IMRT and VMAT planning on 6 MV Flattened and Flattening-Filter-Free Beams of a TrueBeam VirtuaLinac System

Yue Yan

*Department of Medical Physics and Human Oncology, University of Wisconsin-Madison, WI 53792,*

Poonam Yadav

*Department of Human Oncology, University of Wisconsin-Madison and University of Wisconsin, Riverview Cancer Center, Wisconsin Rapids, WI 54494,*

Daniel Saenz and Bhudatt R. Paliwal[*]

*Department of Medical Physics and Human Oncology, University of Wisconsin-Madison, WI 53792,*

**Purpose:** To investigate the accuracy of a TrueBeam VirtuaLinac system for beam commissioning and dose calculation in clinical treatment plans for 6 MV flattened and flattening-filter-free beams.

**Methods:** The TrueBeam VirtuaLinac system was configured to simulate the IAEA phase space data (PSD) outside the gantry head (z=58 cm) for different field sizes. DOSXYZnrc software was used to simulate the water phantom and calculate dose. The beam commissioning data included (1) beam profiles along the x-axis and the diagonal direction, (2) percentage depth dose, (3) output factor. Savitzky-Golay (SG) filters were applied to process the calculated Monte-Carlo (MC) simulation data. MC simulation results were benchmarked with the clinical golden data. Varian Eclipse treatment planning system was commissioned based on the processed MC data (MC plan) and the clinical golden data (control plan). Head & neck and lung cancer cases were used to investigate the differences in dose distribution in both static intensity modulated radiotherapy (IMRT) and volumetric modulated arc therapy (VMAT) plans. Dose volume histograms (DVHs) were used for analysis.

**Results:** For beam profiles, 100% gamma passing rate was obtained for the dose in the *field region*. The maximum averaged deviation in the *penumbra region* was 2.75 mm for a flattened beam of $40 \times 40\ cm^2$ field size. The flatting filter free (FFF) beam with $30 \times 30\ cm^2$ field size at 30 cm depth showed a deviation of 3.06 mm. Maximum deviations (>2%) were observed for the diagonal beam profiles ($40 \times 40\ cm^2$ field size) at 30 cm depth. The MC simulation underestimated the dose in the *tail region* by 3.1% and 2.3% for flattened and FFF beams, respectively. For PDD, the gamma passing rate was near 99%. Large deviations (>2%) were observed at the surface of water phantom. All the deviations for the output factors were within the 2% of threshold value. For clinical IMRT plans (both static IMRT and VMAT), good agreement (<1 Gy in mean and maximum dose) was obtained between the MC plans and the control plans for all organs in the FFF MC plans, and for most of the organs in the flattened MC plans. Large deviations (≥1 Gy) were observed in a lung case for the flattened static IMRT plans in skin and PTV66. For skin, the absolute deviations were 0.78 Gy and 1.28 Gy for mean and maximum doses, respectively. For PTV66, the absolute deviations were 0.16 Gy and 1.25 Gy for mean and maximum doses, respectively.

**Conclusions:** Compared to the flattened beam, the MC plans of the FFF beam showed lower deviations in mean and maximum doses compared to the control plans. Overall, we demonstrated that the beam commissioning using the TrueBeam VirtuaLinac system can provide accurate dose calculation as the clinical golden data.

# 1. INTRODUCTION

In recent decades simulation with Monte-Carlo (MC) techniques has been widely used to estimate the accurate dose distribution for clinical radiotherapy. The simulation of a beam spectrum requires geometrical configuration of the hardware in the head of the machine. However, this information is often not available to the research community due to the commercial interest of the manufacturer. In order to overcome this difficulty, phase space data (PSD) is usually used as a virtual source to simulate the complex hardware. The PSD records the representative pseudo-particles generated during the radiotherapy (RT) process taking into consideration the position, direction, energy and the statistical weight for each particle. The International Atomic Energy Agency (IAEA) promoted a PSD format to facilitate the interchange of phase space files between different Monte Carlo codes[1].

In recent years, Varian released a new class of linear accelerator referred to as the TrueBeam (TB) linear accelerator (linac) system (Varian Medical Systems, Palo Alto, California). In this system, the gantry head is completely redesigned. One of the important characteristics of the TB system is its flattening-filter-free (FFF) operation in the photon beam mode. Compared to the conventional flattened beam, the FFF beam has several advantages, including higher dose rate[2-5], comparatively less external scatter from the gantry head[6-8], lower neutron contamination for high energy beams (>10 MV)[9-11] and improved calculation accuracy due to the removal of the flattening filter[2].

In order to simulate the TB system, Constantin *et al*[12-13] developed a technique to link the computer-aided design (CAD) of the hardware system to the Geant4[14] based MC simulation. The IAEA PSD for the geometry above the jaws of the TB system was released by Varian Medical System. Varian also developed a cloud-based web interface, called "VirtuaLinac". Compared to the self-developed MC model, the VirtuaLinac system has the components below the jaws which are not part of the data package (e.g. the jaws and the baseplate).

Several groups investigated the characteristics of the FFF beam based on the MC simulation technique[15-18]. Belosi *et al* investigated the accuracy of the IAEA PSD for the geometry above the jaws using the geometry of a Clinac 2100 C/D and compared the MC simulation results to measurement data[19]. Parsons *et al* investigated the low-Z target image quality based on the VirtuaLinac system[20]. Gete *et al* compared the differences between the VMAT plans based on MC and Eclipse treatment planning system[21]. The accuracy of the clinical intensity-modulated-radiotherapy (IMRT) plans using beam commissioning data of the VirtuaLinac system is still unclear. In this manuscript, we investigated the accuracy of the PSD in the VirtuaLinac system for beam commissioning and IMRT plans.

## 2.   MATERIALS AND METHODS

### 2.1. MC simulation of the TB system

In our study, the MC simulation of the TB system consisted of two key steps. In step one, the IAEA PSD released by Varian was uploaded to the VirtuaLinac system. We defined the target at Z = 0 and the iso-center of the TB system at Z = 100 cm. The arc scoring plane of the original PSD was at Z=26.7 cm, above the jaws of the TB system. New PSD was calculated using the VirtuaLinac system. The scoring plane of the new PSD was at Z = 58 cm, just outside the gantry head. In the second step, the PSD (Z = 58 cm) was input into the DOSXYZnrc software[22] as a virtual source to simulate the bremsstrahlung photon beam from the gantry head. A water phantom was built based on the DOSXYZnrc MC software.

### 2.2. Important parameters of the MC simulation

For the MC simulation, the cut-off energies for photons and electrons were set as, PCUT=0.01 MeV and ECUT=0.7 MeV, respectively. Non-uniform voxel sizes were used to calculate the 3D dose matrix. For the depths shallower than $d_{max}$, the voxel size was $5\,mm \times 5\,mm \times 3\,mm$. For depth deeper than $d_{max}$, the voxel size was $5\,mm \times 5\,mm \times 1\,cm$. The number of histories (NCASE) to track each particle was $7 \times 10^8$ for simulations of all field sizes. For beam profile and percentage-depth-dose (PDD) calculations, the source-to-surface-distance (SSD) was 100 cm. For the output factor calculations, the SSD was 95 cm and the depth of calculation was 5 cm.

## 2.3. Post processing of the MC simulation data

It is understood that the statistical noise in the MC simulation leads to variations in the calculation results. These variations also influence the choice of the dimension of the dose voxel. If the voxel size is too small (e.g. 1 mm in the in-plane direction), it will lead to a higher calculation imprecision due to the variation in energy deposition from the scattered electrons. Also, the dimension of the dose voxel size is limited by the computer memory. In order to overcome these difficulties, an in-house Fortran code was developed to process the 3D dose matrix of the MC simulation. The Fortran code had three functions. (1) It automatically extracted the data from the 3D dose matrix file generated by the DOSXYZnrc software. (2) Based on the polynomial interpolation algorithm[23], it resampled the beam profile and the PDD of the MC simulation with a fine pixel grid (1 mm in all three directions). (3) It smoothed the MC simulation data based on the 1D Savitzky-Golay (SG) filter[24]. A 2D SG filter using the Matlab[TM] (MathWorks, Natick, MA, USA) was applied to further reduce the variations in the output factor table[25].

## 2.4. Benchmarking the MC simulation

All of the MC simulation results were benchmarked to the standard clinical golden data[26]. Gamma index[27] was used to evaluate the deviation between the MC simulation and the golden data for beam profiles, and PDDs. The reference field sizes were $3 \times 3, 4 \times 4, 6 \times 6, 8 \times 8, 10 \times 10, 20 \times 20, 30 \times 30,$ and $40 \times 40 \ cm^2$ for beam profiles and PDDs. For beam profiles, the reference depths were 1.5, 5, 10, 20, and 30 cm. Three regions were defined for each beam profile and they are summarized as follows:

1. *Field region*: region within 80% of the field size for field sizes $\geq 10 \times 10\ cm^2$ and within 60% of the field sizes for other smaller field sizes ($< 10 \times 10\ cm^2$).

2. *Penumbra region*: off-axis region between 20% and 80% dose lines.

3. *Tail region*: off-axis region below the 20% dose line.

For output factor, the jaws in $x$ and $y$ directions changed from 3 to 40 cm, independently. In total, 81 field sizes were calculated to get the output factor table. Relative deviation ($\delta$) of the output factors between the MC results and the golden data was calculated. $\delta$ is defined as

$$\delta = \frac{|OF_s - OF_{MC}|}{OF_s} \times 100\%. \quad (1)$$

In Eq. (1), $OF_s$ and $OF_{MC}$ refer to output factors of standard clinical golden data and MC simulation, respectively.

**2.5. IMRT plans evaluation**

The TB linac system (Varian Medical Systems, Palo Alto, CA) was commissioned on the Eclipse$^{TM}$ treatment planning system (TPS) based on clinical golden data and the MC simulation data. An Anisotropic-Analytical-Algorithm (AAA) was used to calculate the volume dose distribution for both static IMRT and VMAT plans. The beam energy was 6 MV and dose grid was 2.5 mm for all treatment plans. High definition 120 leaf multi-leaf collimator (MLC) (2.5 mm width in the center and 5 mm width in the peripheral) was used to simulate the treatment plans for all cases.

Head & neck and lung cancer cases were used to investigate the differences in dose distribution in both static IMRT (sliding window) and VMAT plans. These two sites were picked due to their larger treated areas and large number of organs at risk (OAR) requiring complex treatment planning. For the head & neck case (case #1), the dose prescription was 60 Gy in 30 fractions. Two planning-target-volumes (PTVs) were used for plan optimization, PTV60 and PTV 54. The field sizes of the treatment were $16 \times 20\ cm^2$. For the lung case (case #2), the dose prescription was 66 Gy in 33 fractions. The patient had two PTVs,

PTV66 and PTV54. The field sizes were $10 \times 14 \ cm^2$. Both the cases were planned for a uniform dose to the target and a minimum dose to the OARs. For all the treatment plans, the dose in the target was normalized so that 95% of the primary PTV was covered by the 95% isodose line. Dose constraints to OAR's for two clinical cases are summarized in Table 1. Eight treatment plans were generated using the clinical golden data (4 control plans) and the MC simulation data (4 MC plans). For both control and MC plans, it contained flattened and FFF static IMRT plans, and flattened and FFF VMAT plans. The MC plans utilized the same beam fluence and optimization parameters (e.g. arc number, iso-center position, beam angle, etc.) as the corresponding control plans in order to exclude the differences caused by different optimizations. A Matlab code based on the CERR[28] was used to extract the dose distribution and statistically analyze the results from the DICOM data exported from the Eclipse system. Dose evaluation parameters (e.g. mean dose, maximum dose, and organ volume) were benchmarked with the commercial Eclipse treatment planning system. Accurate agreements were obtained for all treatment plans. DVH were used to evaluate the differences between the control plans and the MC plans in terms of target coverage and dose sparing to OAR.

## 3. RESULTS

### 3.1. Beam profiles along $x$ axis

The beam profiles of the 6 MV flattened and the FFF beams were benchmarked with the clinical golden data. The results are shown in Fig.1. We evaluated the deviations between the clinical golden data and the MC simulation in the *field region* and the *penumbra region* (defined above). The gamma index ($\gamma$) analysis[27] in the *field region* is shown in Table 2. The $\gamma$ values were averaged for all reference depths with each field size. The $\gamma$ passing rates were calculated as the fraction of $\gamma$ values which were not greater than 1. For all field sizes, it is evident that both flattened and FFF beams had 100% $\gamma$ passing rate for both $\gamma$ index criteria (2 mm/2% and 3 mm/2%), indicating excellent agreements for the beam profiles in the *field region*.

The spatial absolute deviations ($\Delta d$) in the *penumbra region* between the MC and the clinical golden data are shown in Table 3. The $\Delta d$ was averaged for all reference depths with each field size. Maximum $\Delta d$ for the flattened and FFF beams, was 2.75 mm (for $40 \times 40 \ cm^2$ field size), and 3.05 mm (for $30 \times 30 \ cm^2$ field size), respectively.

**Table 1.** The dose constraints to OARs of case #1 and #2 for the normalized DVH.

| Case #1 | | | |
|---|---|---|---|
| **Organ** | **Volume (%)** | **Dose (Gy)** | **Maximum Dose (Gy)** |
| Left eye | N/A | N/A | 4 |
| Right eye | N/A | N/A | 4 |
| Cord+3 mm | N/A | N/A | 43 |
| Normal | N/A | N/A | 43 |
| Brain | N/A | N/A | 45 |
| Right shoulder | N/A | N/A | 5 |
| Right submandibular | N/A | N/A | 26 |
| Case #2 | | | |
| Cord expand | N/A | N/A | 40 |
| Bilateral lungs-PTV | 25 | 17.5 | N/A |
| | 10 | 30 | N/A |
| Heart | 40 | 30 | N/A |
| Esophagus | 40 | 40 | N/A |
| Normal | N/A | N/A | 30 |

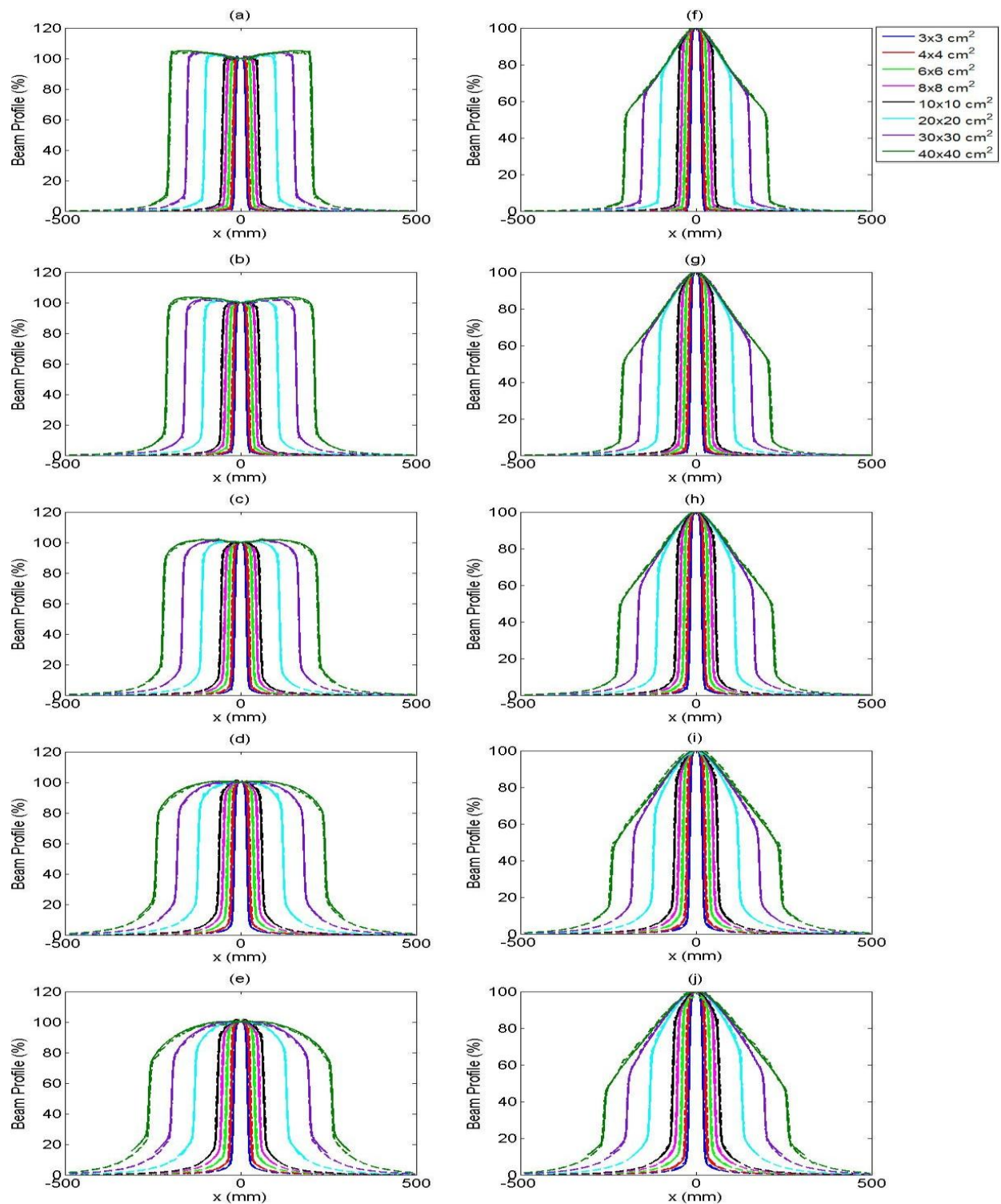

**Figure 1.** Beam profile for the 6 MV flattened beam (left column) and 6 MV FFF beam (right column). . The reference depths are 1.5 cm ((a) and (f)), 5 cm ((b) and (g)), 10 cm ((c) and (h)), 20 cm ((d) and (i)), 30 cm ((e) and (j)). The solid lines represent the clinical golden data and the dashed lines represent the MC simulation data.

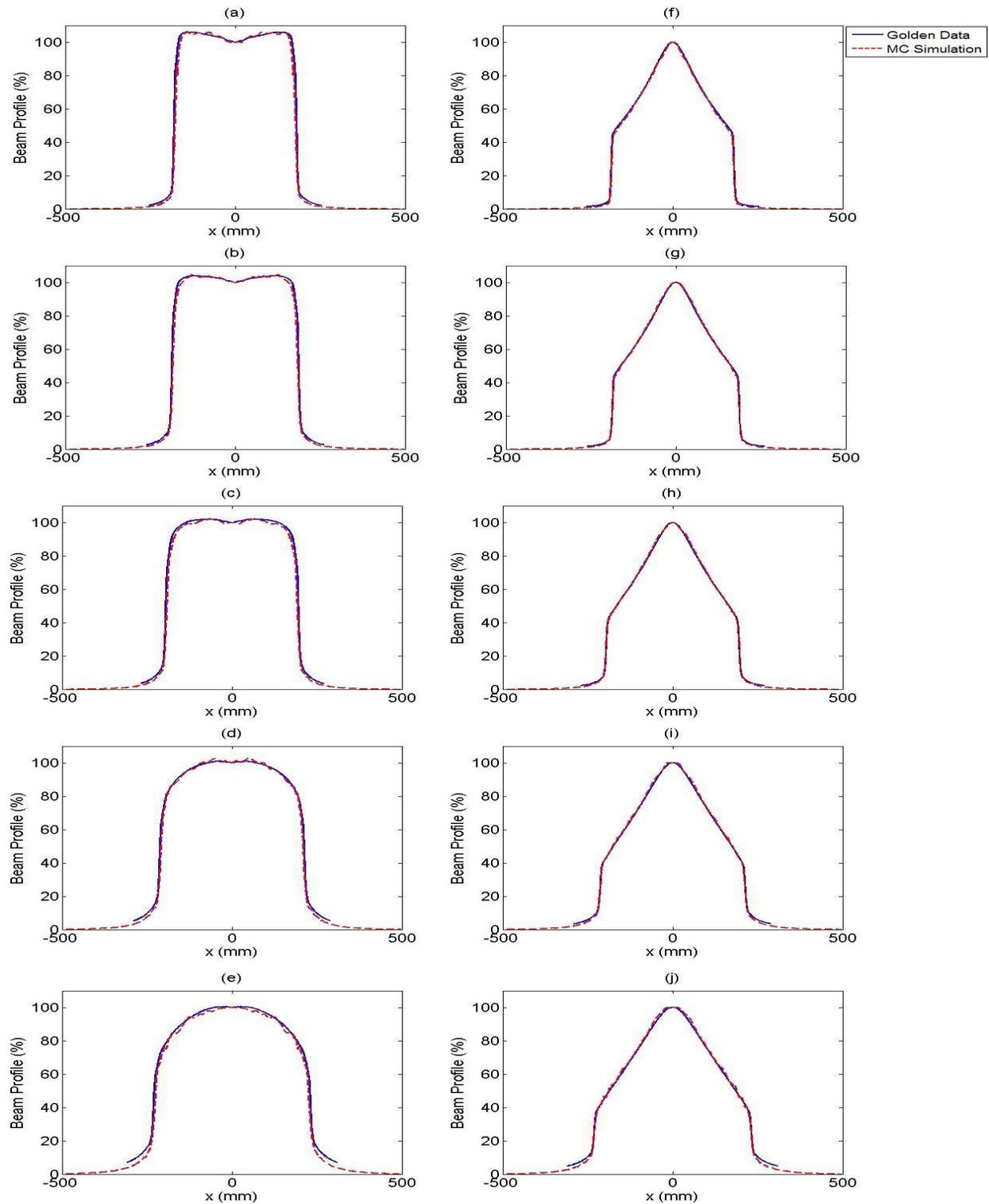

**Figure 2.** Diagonal beam profile for the 6 MV flattened beam (left column) and 6 MV FFF beam (right column). .The field size is $40 \times 40\ cm^2$. The reference depths are 1.5 cm ((a) and (f)), 5 cm ((b) and (g)), 10 cm ((c) and (h)), 20 cm ((d) ,and (i)), 30 cm ((e), and (j)). The blue solid lines represent the clinical golden data and the red dashed lines represent the MC simulation data.

### 3.2. Beam profiles along the diagonal direction

The beam profiles along the diagonal direction ($40 \times 40 \ cm^2$ field size) are shown in Fig. 2 for flattened and FFF beams. The $\gamma$ index for the *field region* and the deviation in the *penumbra region* ($\Delta d_{diag}$) are shown in Table 4. For the two selected criteria, the $\gamma$ index of both flattened and FFF beams reached a 100% passing rate. In the analysis of penumbra, the averaged $\Delta d_{diag}$ were 1.96 mm and 2.14 mm for flattened and FFF beams, respectively. Good agreements were obtained between the MC simulation and the golden data for most of the results. Deviations were observed in the *tail region* of the beam profiles for $40 \times 40 \ cm^2$ at $30 \ cm$ depth for both flattened and FFF beams. The maximum dose deviations were 3.1% and 2.3% for flattened and FFF beams, respectively. Based on previous discussion and literature review[13], the deviations in the *tail region* may be due to the statistical imprecision in the $x$ and $y$ coordinates of the electron PSD generated by the Parmela simulation of the TB hardware system.

### 3.3. PDD

MC simulations of the PDDs of selected field sizes are shown in Fig. 3 for both flattened and FFF beams. The $\gamma$ index analysis is shown in Table 5. For all reference field sizes, the $\gamma$ index passing rates were around 99%. For most of data point, the maximum deviations between the MC simulation and the golden data were less than 1%. For the dose at the surface of the water phantom, relatively large deviations (>2%) were obtained between the MC results and the golden data. This is due to the steep gradient of the PDD curve in the build-up region at the superficial depth of the water phantom. .

### 3.4. Output factor

The relative deviations ($\delta$) of the output factors for 81 field sizes of flattened and FFF beams are shown in Fig. 4. These deviations were within 2% for all field sizes. Large deviations (>1.5%) were observed for large asymmetric field sizes (e.g. $40 \times 3 \ cm^2$).

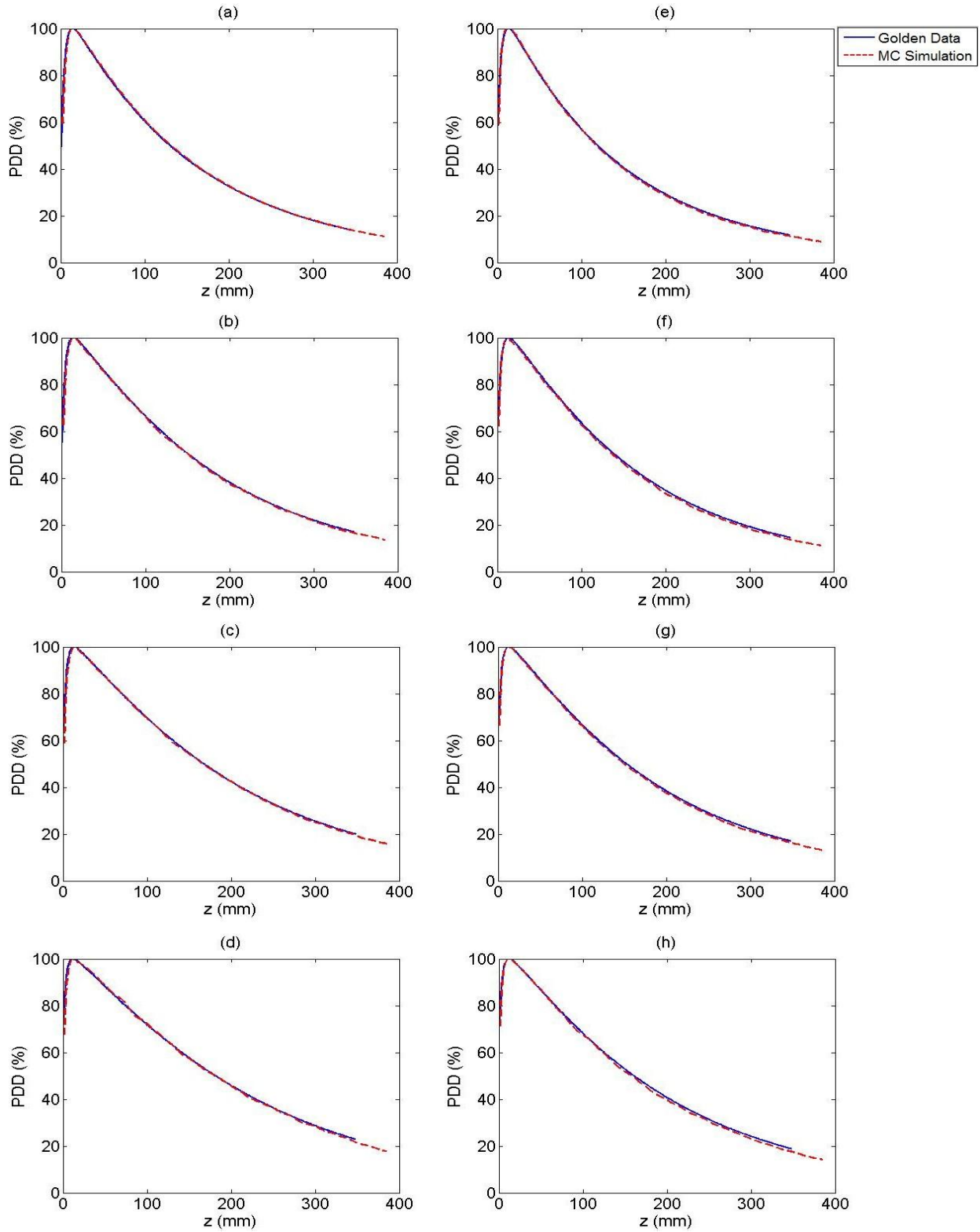

**Figure 3.** PDD of the 6 MV flattened (left column) and FFF (right column) beams. The SSD is 100 $cm$. The field sizes are: 3 × 3 $cm^2$ ((a) and (e)), 10 × 10 $cm^2$ ((b) and (f)), 20 × 20 $cm^2$ ((c) and (g)), and 40 × 40 $cm^2$ ((d) and (h)).

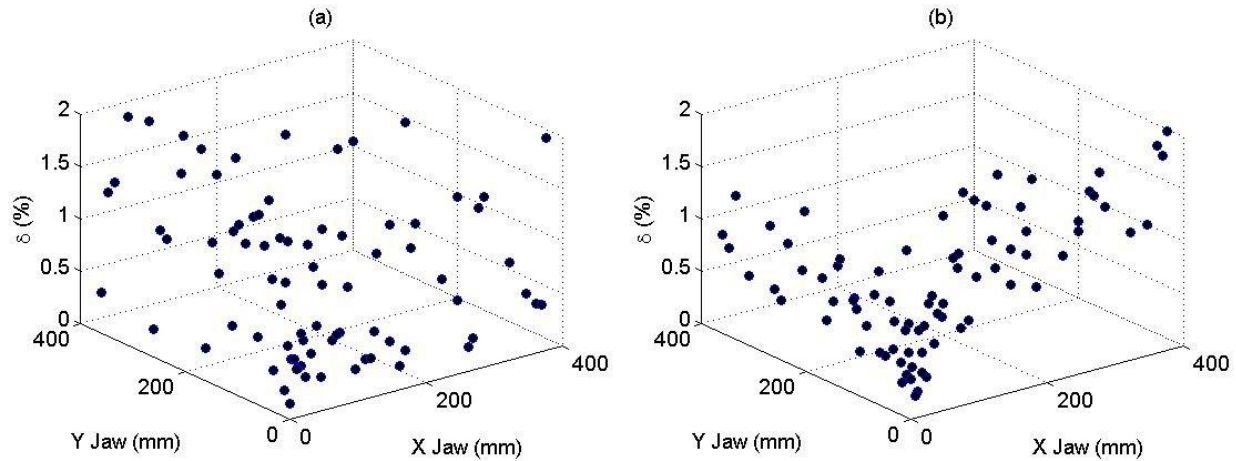

**Figure 4.** The relative deviations of the output factors for the 6 MV (a) flattened, and (b) FFF beam between the MC simulation and the clinical golden data.

**Table 2.** γ index analysis in the field region of the beam profiles for the 6 MV flattened and the FFF beams.

| | γ Index of the Beam Profile | | | | | | |
|---|---|---|---|---|---|---|---|
| | **6 MV Flattened Beam** | | | | **6 MV FFF Beam** | | |
| | **2 mm/2%** | | **3 mm/2%** | | **2 mm/2%** | | **3 mm/2%** |
| **Field size (cm²)** | **γ** | **Passing Rate (%)** | **Γ** | **Passing Rate (%)** | **γ** | **Passing Rate (%)** | **γ** | **Passing Rate (%)** |
| **3x3** | 0.10 | 100.0 | 0.07 | 100.0 | 0.11 | 100.0 | 0.07 | 100.0 |
| **4x4** | 0.17 | 100.0 | 0.11 | 100.0 | 0.17 | 100.0 | 0.11 | 100.0 |
| **6x6** | 0.12 | 100.0 | 0.08 | 100.0 | 0.12 | 100.0 | 0.08 | 100.0 |
| **8x8** | 0.17 | 100.0 | 0.11 | 100.0 | 0.17 | 100.0 | 0.12 | 100.0 |
| **10x10** | 0.15 | 100.0 | 0.12 | 100.0 | 0.14 | 100.0 | 0.10 | 100.0 |
| **20x20** | 0.15 | 100.0 | 0.11 | 100.0 | 0.14 | 100.0 | 0.10 | 100.0 |
| **30x30** | 0.16 | 100.0 | 0.12 | 100.0 | 0.14 | 100.0 | 0.10 | 100.0 |
| **40x40** | 0.18 | 100.0 | 0.14 | 100.0 | 0.18 | 100.0 | 0.15 | 100.0 |

**Table 3.** The absolute penumbra deviation between the golden data and the MC simulation.

| | $\Delta d$ (mm) | |
|---|---|---|
| **Field Size (cm$^2$)** | **6 MV Flattened Beam** | **6 MV FFF Beam** |
| **3x3** | 0.48 | 0.62 |
| **4x4** | 0.36 | 0.39 |
| **6x6** | 0.46 | 0.48 |
| **8x8** | 0.75 | 0.66 |
| **10x10** | 0.60 | 0.54 |
| **20x20** | 1.19 | 2.59 |
| **30x30** | 1.36 | 3.05 |
| **40x40** | 2.75 | 1.63 |

**Table 4**. γ index analysis in the *field region* and absolute penumbra deviations of the diagonal beam profiles for the 6 MV flattened and FFF beam. The field size is $40 \times 40\ cm^2$.

| γ Index of the Diagonal Beam Profile | | | | | | | |
|---|---|---|---|---|---|---|---|
| **6 MV Flattened Beam** | | | | **6 MV FFF Beam** | | | |
| 2 mm/2% | | 3 mm/2% | | 2 mm/2% | | 3 mm/2% | |
| γ | Passing Rate (%) | γ | Passing Rate (%) | γ | Passing Rate (%) | γ | Passing Rate (%) |
| 0.2 | 100 | 0.19 | 100 | 0.21 | 100 | 0.18 | 100 |
| $\Delta d_{diag}$ (mm) | | | | | | | |
| **6 MV Flattened Beam** | | | | **6 MV FFF Beam** | | | |
| 1.96 | | | | 2.14 | | | |

**Table 5.** Gamma index analysis of the PDD for the 6 MV flattened and FFF beam.

| | | γ index of the PDD | | | | | | |
|---|---|---|---|---|---|---|---|---|
| | | 6 MV Flattened Beam | | | | 6 MV FFF Beam | | |
| | | 2 mm/2% | | 3 mm/2% | | 2 mm/2% | | 3 mm/2% |
| Field size (cm²) | γ | Passing Rate (%) | Γ | Passing Rate (%) | γ | Passing Rate (%) | γ | Passing Rate (%) |
| 3x3 | 0.23 | 99.0 | 0.20 | 99.0 | 0.32 | 99.0 | 0.29 | 99.0 |
| 4x4 | 0.26 | 98.7 | 0.22 | 98.9 | 0.32 | 99.3 | 0.29 | 99.3 |
| 6x6 | 0.30 | 99.1 | 0.26 | 99.1 | 0.43 | 98.9 | 0.40 | 98.9 |
| 8x8 | 0.21 | 98.9 | 0.17 | 99.0 | 0.40 | 99.1 | 0.37 | 99.1 |
| 10x10 | 0.29 | 99.0 | 0.26 | 99.1 | 0.45 | 99.6 | 0.42 | 99.6 |
| 20x20 | 0.26 | 99.3 | 0.23 | 99.3 | 0.39 | 99.1 | 0.36 | 99.1 |
| 30x30 | 0.32 | 99.4 | 0.29 | 99.4 | 0.40 | 99.3 | 0.37 | 99.4 |
| 40x40 | 0.29 | 99.3 | 0.26 | 99.4 | 0.45 | 99.3 | 0.42 | 99.4 |

### 3.5. Clinical IMRT plans' evaluation

DVHs of control and MC plans are shown in Fig.6 and Fig.7. The MC plans are in good agreement with the control plans for both static IMRT and VMAT plans. For both cases, the FFF beam MC plans provided accurate simulation as the control plans (<0.5 Gy for mean dose and maximum dose) for all organs for both static IMRT and VMAT plans. Compared to the FFF beam MC plans, the flattened beam MC plans showed relatively larger deviations from the control plans. In case #1, for the flattened beam plans, as shown in Fig.6 (a) and (c), the skin had the maximum deviation in the static IMRT plan. The absolute deviations in mean and maximum dose for skin reached up to 0.34 Gy and 0.61 Gy, respectively. In case #2, for flattened beam plans, as shown in Fig. 7 (a), relatively important differences between the MC and the control plans were observed in the static IMRT plans for skin and PTV66. For skin, the deviations were 0.78 Gy and 1.28 Gy for mean dose and maximum dose, respectively. For PTV66, the deviations were 0.16 Gy and 1.25 Gy for mean and maximum dose, respectively. For all other organs in Fig. 7 (a), the deviations were less than 0.5 Gy for both mean and maximum doses. For VMAT plans, as shown in Fig.7 (c), the larynx had the maximum dose deviation. The difference was 0.36 Gy for both mean and maximum doses. Overall, no clinically significant differences were noticed.

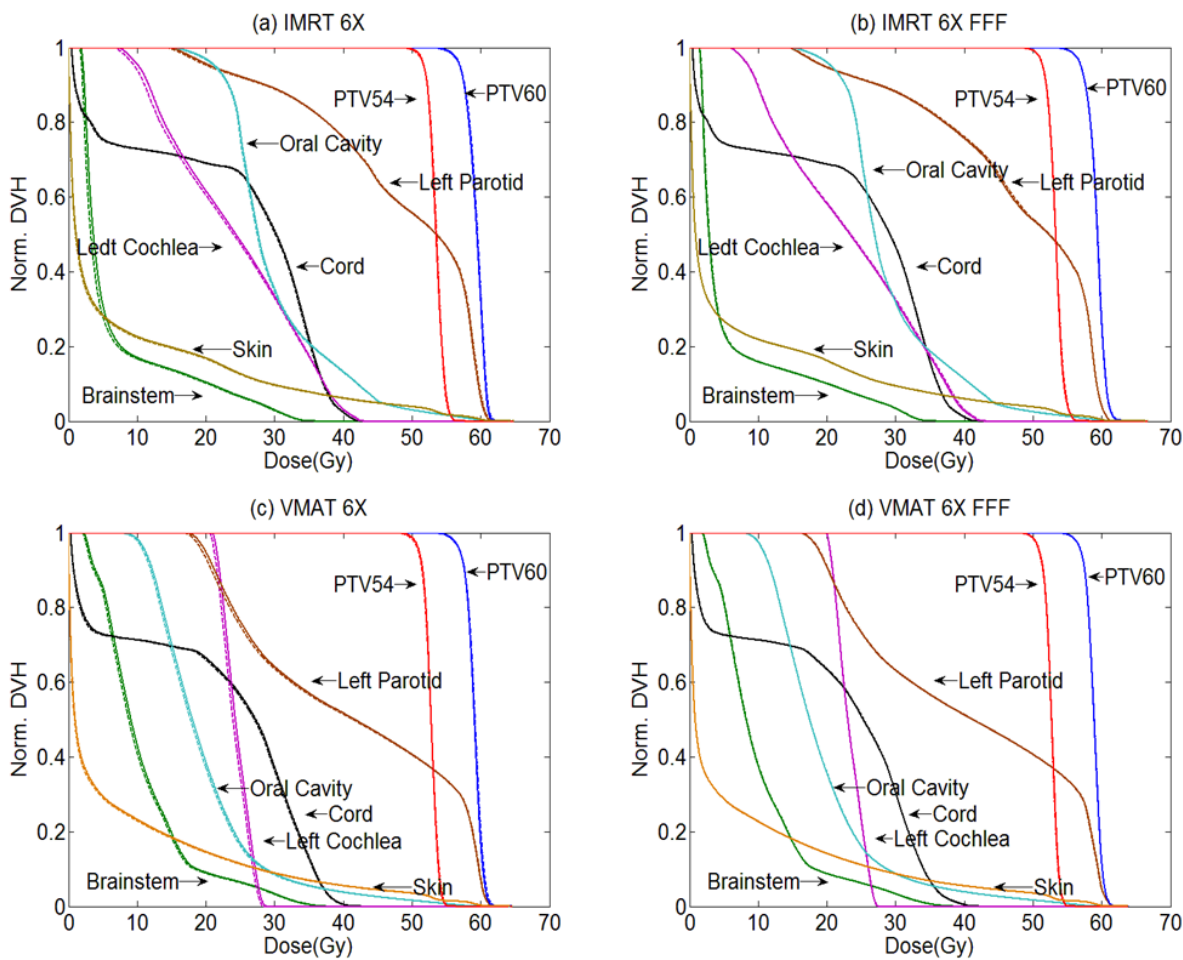

**Figure 6.** DVH of case #1 (head & neck case) of (a) flattened static IMRT, (b) FFF static IMRT, (c) flattened VMAT, and (d) FFF VMAT plans. The solid lines represent the control plans and the dashed lines represent the MC plans.

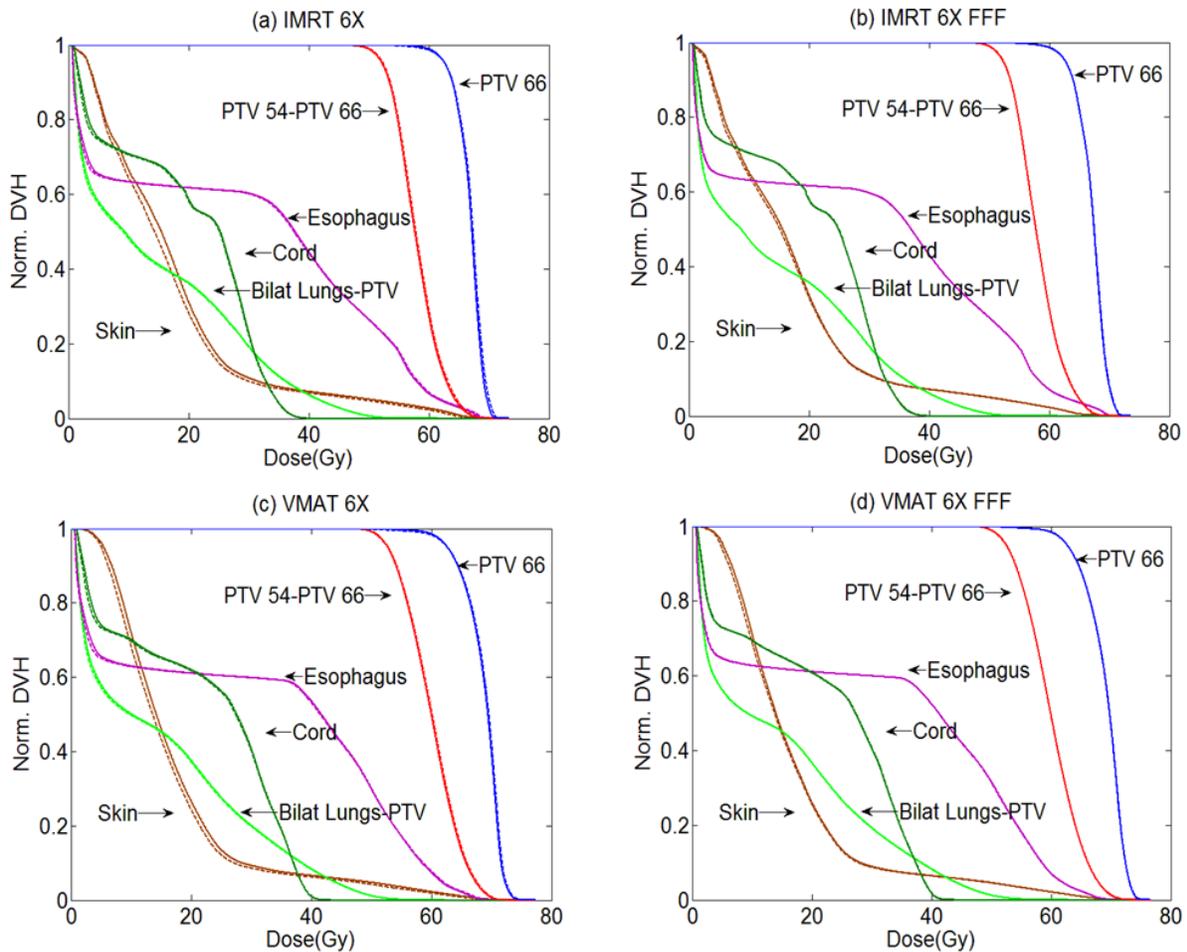

**Figure 7.** DVH of case #2 (lung case) of (a) flattened static IMRT, (b) FFF static IMRT, (c) flattened VMAT, and (d) FFF VMAT plans. The solid lines represent the control plans and the dashed lines represent the MC plans.

## 4. DISCUSSION

Based on Fig.6 and Fig.7, the FFF beam contributed to improved dose calculation accuracy in the clinical IMRT plans compared to the flattened beam. The higher deviations between the MC plans and the control plans for the flattened beam can possibly be explained by the higher statistical imprecisions in the output factor for the flattened beam. Based on Fig.4, compared to the FFF beam, the output factor of the flattened beam shows larger deviations to the clinical golden data. The mean value of $\delta$ for the FFF beam is 0.74%. For the flattened beam, this value reaches 0.80%. In addition, the FFF beam shows lower variations in the output factor compared to the flattened beam[2]. These two effects possibly contribute to the improved dose calculation accuracy of the FFF beam in clinical treatment plans compared to the flattened beam. .

It is easy to modify and design new hardware components of the TB system based on the VirtuaLinac MC model. How accurate is the MC simulation in the clinical treatment plan calculation is an open question. In this manuscript, we evaluated the accuracy of the entire chain of phase space files generated by the VirtuaLinac system in clinical plans. In addition, beam commissioning data of the VirtuaLinac system may be used to supplement directly measured data. The relative magnitude of agreement between MC and golden beam data presented in this research may assist a physicist in terms of the percent dose deviation that one may expect when using MC data for verification purposes.

## 5. CONCLUSIONS

In this manuscript, we investigated the accuracy of the 6 MV flattened and FFF beams of the VirtuaLinac system in terms of beam commissioning and clinical IMRT treatment plans. Good agreement was obtained between the MC simulation and the clinical golden data for both flattened and FFF beams. For beam commissioning, the maximum deviations between the MC simulation and the golden data were observed in the diagonal beam profile ($40 \times 40 \ cm^2$ field size) at 30 cm depth. The MC simulation underestimated the dose in the *tail region* of the beam profile up to 3.1% and 2.3% for flattened and FFF beams, respectively. For clinical applications, the MC FFF beam plans showed lower deviations to the control plans compared to the MC flattened beam plans. These deviations are not clinically significant and hence demonstrate the acceptability of the VirtuaLinac system. Based on our finding we believe it can be used for other treatment sites successfully. Overall, the presented study demonstrated that dose calculation using the beam commissioning data of the VirtuaLinac system can provide accurate simulations for the clinical IMRT (including static IMRT and VMAT) treatment plan calculations.

**ACKNOWLEDGEMENTS**

The authors in this manuscript would like to thank Dr. Willie Shaw from the Department of Medical Physics, University of the Free State, for sharing part of his Fortran codes to read the data from the 3D data files generated by the DOSXYZnrc MC software. They would also like to thank Dr. Daren Sawkey, from the Varian, for his technical support of the VirtuaLinac system.

[*]Author to whom correspondence should be addressed. Electronic addresses: paliwal@humonc.wisc.edu. Telephone: 608 263 8514.